\documentstyle[12pt]{article}
\input psfig
\long\def\comment#1{}
\begin{document}
\title{The Cyclic Universe: Some Historical Notes}
\author{Subhash Kak\\
%Department of Electrical \& Computer Engineering\\
Louisiana State University\\
Baton Rouge, LA 70803-5901\\}
%Email: {\tt kak@ee.lsu.edu}}
%\date{{\bf {\large Technical Report ECE-01-04}\\
%January 19, 2001}}
\maketitle

\begin{abstract}
The cyclic model of the universe has an old history in India.
It has held the pre\"{e}minent position 
regarding the origins of the universe, and it 
is described in astronomical texts, Pur\={a}\d{n}ic encyclopaedias, and
philosophical literature.
Within its current cycle, which is supposed to have begun several
billion
years ago, are smaller cycles of
pole changes and extinctions on earth.
Salient features of
the cyclic universe model are presented here.

%{\bf Keywords:} Inflationary universe, big bang, speed of light\\

%{\bf PACS no. 04.20}
\end{abstract} 
 
\thispagestyle{empty}

\subsection*{Introduction}

With the renewed interest in the cyclic\cite{St02} and
the quasi-steady state cosmo-\ logical\cite{Ho00,Na02}  models,
it seems appropriate to look at the early history of such 
ideas.
In this paper, we look at the conception of cyclic universes
in the Indic tradition.
The development of this conception in India owes
to three fundamental notions that are come across in
the earliest texts: (i), time is endless and space has
infinite extension; (ii), earth is not the center of
the universe; and (iii), laws govern all development,
including the creation and destruction of the universe.

Astronomy was a very highly valued science in India\cite{Ka00a,Ka00b}.
It was believed that there were connections between
the physical and the psychological worlds, and an equivalence
between
the outer cosmos and the inner cosmos of the individual.
This is expressed 
in the famous sentence; {\it yat pin\d{d}e tad brahm\={a}\d{n}\d{d}e}, 
``as in the cell so in the universe". 
This belief may have contributed to a generalization from the
cyclic nature of human life into a corresponding theory
about the cosmos.

Since thought
and speculation have played a large role in the development
of Indic ideas, the cyclic model may be taken to provide the 
philosopher's conception of the universe, complementing
the other ten conceptions that Fred Hoyle described in
his popular book, {\it Ten Faces of the Universe}\cite{Ho77}.

The Indian cyclic model assumes the existence of 
countless
island universes, which go through their own periods of development
and destruction.
The conception of cyclicity is taken to be recursive.

For an early 
exposition of Indic astronomical and cosmological
ideas, one may like to read al-B\={\i}r\={u}n\={\i}'s
classic history of Indian science, composed in 
1030 A.D.\cite{Sa89}, keeping in mind
that
al-B\={\i}r\={u}n\={\i} did not correctly 
understand all Indian material;
for an even earlier, popular, view of Indian ideas, see
the Yoga V\={a}si\d{s}\d{t}ha\cite{Ve93,Ka99};
for a review of
the birth and development of early
Indian astronomy, see\cite{Ka00a,Ka00b};
for a broad overview, see\cite{Ka01a}.

\paragraph{1}
%\subsection{The Cycle Periods}
{\it Cycle periods.}
The aspect of the Indian cyclic model that astrophysicists 
are most familiar with is a cycle called
the {\it kalpa}, the ``day'' of Brahm\={a},
which is 4.32 billion years long.
There is an equally long ``night'', and 360 such ``days''
and ``nights'' constitute a year of Brahm\={a}. The life
of Brahm\={a} is 100 such years, which is
thus 311,040 billion years long. 

Then there exist longer periods in an endless process. 
Al-Biruni\cite{Sa89} lists longer
periods going to 
$10,782,449,978,758,523,781,120 \times 10^{27}$
  kalpas, which is called
{\it tru\d{t}i}. 

\paragraph{2}
{\it Sub-periods.}
Within the {\it kalpa} are fourteen cycles of local creation and
destruction, called {\it manvantaras}, each lasting 
306.720 million years.
Each of these periods is to be taken to be a period of local destruction 
(extinction) and
subsequent regeneration on
the earth.
Within each manvantara are 71 smaller cycles, called
{\it mah\={a}yugas}.

These periods are described by \={A}ryabha\d{t}a, the great
astronomer born in 476 A.D.,
 in his \={A}ryabha\d{t}\={\i}ya\cite{Sh76}.
As is well known,
\={A}ryabha\d{t}a
presented  the rotation information of the outer
planets with respect
to the sun,
meaning that his system
was partially heliocentric;
furthermore, he considered the earth to be
rotating on its own axis.

In another version, described in the S\={u}ryasiddh\={a}nta
and 
the Pur\={a}\d{n}as, each kalpa equals 1,000 yugas of
4.32 million years.

Yoga V\={a}si\d{s}\d{t}ha 6.1.22\cite{Ve93}
says that directions depend on the position of the poles,
the movements of the stars, the sun, and the moon, and
that the directions change from one sub-period to another.
\paragraph{3}

{\it Ideas related to time.}
Al-B\={i}r\={u}n\={\i} 
says this about India ideas on time\cite{Sa89}:
\begin{quote}
The Hindus have divided duration into two periods,
a period of {\em motion}, which has been determined
as {\em time}, and a period of {\em rest}, which
can only be determined in an imaginary way according
to the analogy of that which has first been
determined, the period of motion.
The Hindus hold the eternity of the Creator to be
{\em determinable}, not {\em measurable}, since it
is infinite.

They do not, by the word {\em creation}, understand
a {\em formation of something out of nothing.}
They mean by creation only the working with a piece of
clay, working out various combinations and figures in it,
and making such arrangements with it as will lead to
certain ends and aims which are potentially in it
(page 321, vol. 1).
\end{quote}

\paragraph{4}
{\it Island-universes.}
The universe is split up into many island-universes,
as described in the Yoga Vasisththa 6.2.59\cite{Ve93}:

\begin{quote}
I saw countless creations though they did not know of one
another's existence.
Some were coming into being, others were perishing, all
of them had different shielding atmospheres (from five to
thirty-six atmospheres).
There were different elements in each, they were inhabited
by different types of beings in different stages of evolution..
[In] some there was apparent natural order in others there
was utter disorder, in some there was no light and
hence no time-sense. 
\end{quote}

\paragraph{5}

{\it Evolution.}
The idea of evolution is basic to all Indian thought.
The Indian theory of evolution, which is supposed to apply
both to the individual and the cosmos, is called S\={a}\d{m}khya.
In it,
the basic entities are pure consciousness and materiality (nature).
Nature has three constituent qualities
({\it gu\d{n}as}) called {\it sattva, rajas}, and {\it tamas}, and
as the balance between these three changes the universe evolves.

Out of the interplay of the five basic elements arise 
other principles ({\it tattva}):
five subtle elements, five action senses, five senses of
perception, mind, egoity, and intellect.
The evolutionary sequence goes through many levels.
The tattvas help in the
the emergence of life out of inert matter. 
The gu\d{n}as are not to be taken as abstract principles
alone. Indian thought believes that structure in
nature is recursive, and the gu\d{n}as show up in various 
forms at different levels of expression.

The texts imply that ingredients for the growth of
life are available throughout the universe. Infinite
number of universes are conceived, so each new one
is created like a bubble in an ocean of bubbles. 
Indian evolution theory is like the neutral theory. If
the gene function is seen through the agency of the
three gunas, then evolution has a net genetic drift.
The tattvas are not discrete and their varying
expression creates the diversity of life in and
across leading different species. Each sensory
and motor tattva is mapped into a corresponding organ.
Indian thought conceives of 8.4 million species, which
number
is impressive, considering that modern authorities (such
as Graur and Li in their ``Fundamentals of Molecular 
Evolution", page 436) estimate the number of extant 
species to be 4.5 - 10 million.

Schr\"{o}dinger
thought\cite{Mo89} that 
the S\={a}\d{m}khyan tattvas were the most plausible model for 
the evolution of the sensory organs.

 A quote on evolution on earth from the Yoga Vasishtha (6.1.21)\cite{Ve93}:
\begin{quote}
 I remember that once upon a time there was nothing on
  this Earth, neither trees and plants, nor even mountains. 
  For a period of eleven thousand years (4 million Earth 
  years) the Earth was under lava... [Later] apart from 
  the polar region, the rest of the Earth was covered by
  water. And then forests enveloped the Earth, and great 
  asuras (demons) ruled. 

  Then there arose great mountains, but without any human 
  inhabitants. For a period of ten thousand years (almost 4 
  million Earth years) the Earth was covered with the corpses 
  of the asuras (daityas)."

\end{quote}

  Indicating the presence of other animals while the giant
  asuras were on Earth, YV suggests that man arose later.
  YV also speaks of minor ages of destruction on Earth that
  correspond to the yugas.

\paragraph{6}
{\it Size of island-universe.}
It appears that the
size of each island universe was taken to be
equal to the distance that light would travel in 
twenty-four hours.
The speed of light is taken in Indian texts is said
to be 4,404 yojanas per nime\d{s}a, which is
almost exactly 186,000 miles per second\cite{Ka98}!
The earliest mention of this is in texts that are
more than six hundred years old.

To have chanced upon the correct number for speed of light, before
it could have possibly been measured, is the most amazing 
coincidence in all of science.
This coincidence is much more striking than the general
correctness of the age of life on earth in the Indic view,
and much more than the coincidences that are a part
of the process of scientific discovery\cite{Ba86}.

\paragraph{7}
{\it Evolution of life.}
The Indians believed that all life can be divided
into three  classes (Ch\={a}ndogya Upani\d{s}ad 6.3.1): 
``In truth, beings have here three kinds of seeds, born
from the egg, born alive, and born from the germ.''
Given that it is also affirmed that life on other
planets exists and that there
was a gradual rise of life on the earth, it
would appear that this implied a belief in a
panspermia theory.

\paragraph{8}
{\it Inner cosmos.}
In Indian mythology, the continents are in concentric circles.
Wrongly applied to the
outer cosmos, 
this Pur\={a}\d{n}ic cosmology represents the inner cosmos of
the individual on a scale that equals the size of
the universe. It, therefore, brings in outer
astronomy only in an incidental fashion. The earth
of the Pur\={a}\d{n}as is the individual pictured as the plane
that touches the navel.  Below the navel are the underworlds;
above the navel are the sun and the moon in the head,
and beyond them the planets and the stars.
The perceived dimensions of the sun and the moon in
this conception relate to the inner cosmos.

\paragraph{9}
{\it Another cyclic model.}
I have developed
a cosmological model inspired by S\={a}\d{m}khyan
ideas\cite{Ka01b}.
Here two phases collide, where one collapses the
quantum state of the other.

\subsection*{Conclusions}
This brief note is an introduction to Indic cosmological 
ideas on a cyclic universe.


\begin{thebibliography}{10}

\bibitem{Ba86}
J.D. Barrow and F.J. Tipler, {\it The Anthropic Cosmological Principle.}
(Oxford University Press, New York, 1986).

\bibitem{Ho77}
F. Hoyle, {\it Ten Faces of the Universe.}
W.H. Freeman, San Francisco.

\bibitem{Ho00}
F. Hoyle, G. Burbidge, and J.V. Narlikar,
{\it A Different Approach to Cosmology.}
Cambridge University Press, 2000.

\bibitem{Ka98}
S. Kak, 
``The Speed of Light and Puranic Cosmology,''
{\it  Annals of the Bhandarkar Oriental 
Research Institute}, vol. 80, 1999, pp. 113-123. 
LANL preprint physics/9804020.


\bibitem{Ka99}
S. Kak, 
``Concepts of Space, Time, and Consciousness in Ancient India.''
LANL preprint physics/9903010.


\bibitem{Ka00a}
S. Kak, 
``Birth and Early Development of Indian Astronomy.''
In {\it Astronomy Across Cultures: The History of Non-Western Astronomy,}
Helaine Selin (ed), Kluwer, 2000, pp. 303-340.
LANL preprint physics/0101063.

\bibitem{Ka00b}
S. Kak, 
{\it The Astronomical Code of the \d{R}gveda.}
Munshiram Manoharlal, New Delhi, 2000.


\bibitem{Ka01a}
S. Kak, 
{\it The Wishing Tree.}
Munshiram Manoharlal, New Delhi, 2001.


\bibitem{Ka01b}
S. Kak, 
``Collapse, expansion, and a variable speed of light.''
LANL preprint astro-ph/0101455.

\bibitem{Mo89}
W. Moore, {\it Schr\"{o}dinger: Life and Thought.} Cambridge
University Press, Cambridge, 1989.


\bibitem{Na02}
J.V. Narlikar, R.G. Vishwakarma, and G. Burbidge,
``Interpretations of the accelerating universe.''
LANL Archive astro-ph/0205064.

\bibitem{Sa89}
E.C. Sachau, {\it Alberuni's India.} Delhi, 1989 [1910].

\bibitem{Sh76}
K.S. Shukla and K.V. Sarma, {\em \={A}ryabha\d{t}\={\i}ya of
\={A}ryabha\d{t}a.}
Indian National Science Academy, New Delhi, 1976.




\bibitem{St02}
P.J. Steinhardt and N. Turok, ``The cyclic universe: an informal
introduction.'' LANL Archive astro-ph/0204479.

\bibitem{Ve93}
S. Venkatesananda, (tr.), {\it V\={a}si\d{s}\d{t}ha's Yoga.}
State University of New York Press, Albany, 1993. For the Sanskrit text, see
{\it Yoga V\={a}si\d{s}\d{t}ha}, 1981. Munshiram Manoharlal, Delhi.





\end{thebibliography}
\end{document}